\documentclass{article}
\usepackage[utf8]{inputenc}
\usepackage{geometry}
\usepackage{subcaption}
\usepackage{hyperref, url}
\usepackage{setspace}
\usepackage{titling}
\usepackage{authblk}

\title{Record high solar irradiance in Western Europe during first COVID-19 lockdown largely due to unusual weather}

\author[1]{Chiel C. van Heerwaarden \texttt{(chiel.vanheerwaarden@wur.nl)}}
\author[1]{Wouter B. Mol}
\author[1]{Menno A. Veerman}
\author[1]{Imme B. Benedict}
\author[1]{Bert G. Heusinkveld}
\author[2]{Wouter H. Knap}
\author[3]{Stelios Kazadzis}
\author[3]{Natalia Kouremeti}
\author[4,5]{Stephanie Fiedler}
\affil[1]{Meteorology and Air Quality Group, Wageningen University, Wageningen, The Netherlands}
\affil[2]{Royal Netherlands Meteorological Institute, De Bilt, The Netherlands}
\affil[3]{Physikalisch-Meteorologisches Observatorium Davos, World Radiation Center (PMOD-WRC), Davos, Switzerland}
\affil[4]{University of Cologne, Institute of Geophysics and Meteorology, Cologne, Germany}
\affil[5]{Hans-Ertel-Centre for Weather Research, Climate Monitoring and Diagnostics, Bonn/Cologne, Germany}

\date{January 21, 2021}

\usepackage{graphicx}

\begin{document}

\maketitle

\section{Abstract}
Spring 2020 broke sunshine duration records across Western Europe.
The Netherlands recorded the highest surface irradiance since 1928, exceeding the previous extreme of 2011 by 13\,\%, and the diffuse fraction of the irradiance measured a record low percentage (38\,\%).
The coinciding irradiance extreme and a reduction in anthropogenic pollution due to COVID-19 measures triggered the hypothesis that cleaner-than-usual air contributed to the record.
Based on analyses of ground-based and satellite observations and experiments with a radiative transfer model, we estimate a 1.3\,\% (2.3 W m$^{-2}$) increase in surface irradiance with respect to the 2010-2019 mean due to a low median aerosol optical depth, and a 17.6\,\% (30.7 W m$^{-2}$) increase due to several exceptionally dry days and a very low cloud fraction overall.
Our analyses show that the reduced aerosols and contrails due to the COVID-19 measures are far less important in the irradiance record than the dry and particularly cloud-free weather.

\begin{figure}[ht]
\centering
\includegraphics[width=1\textwidth]{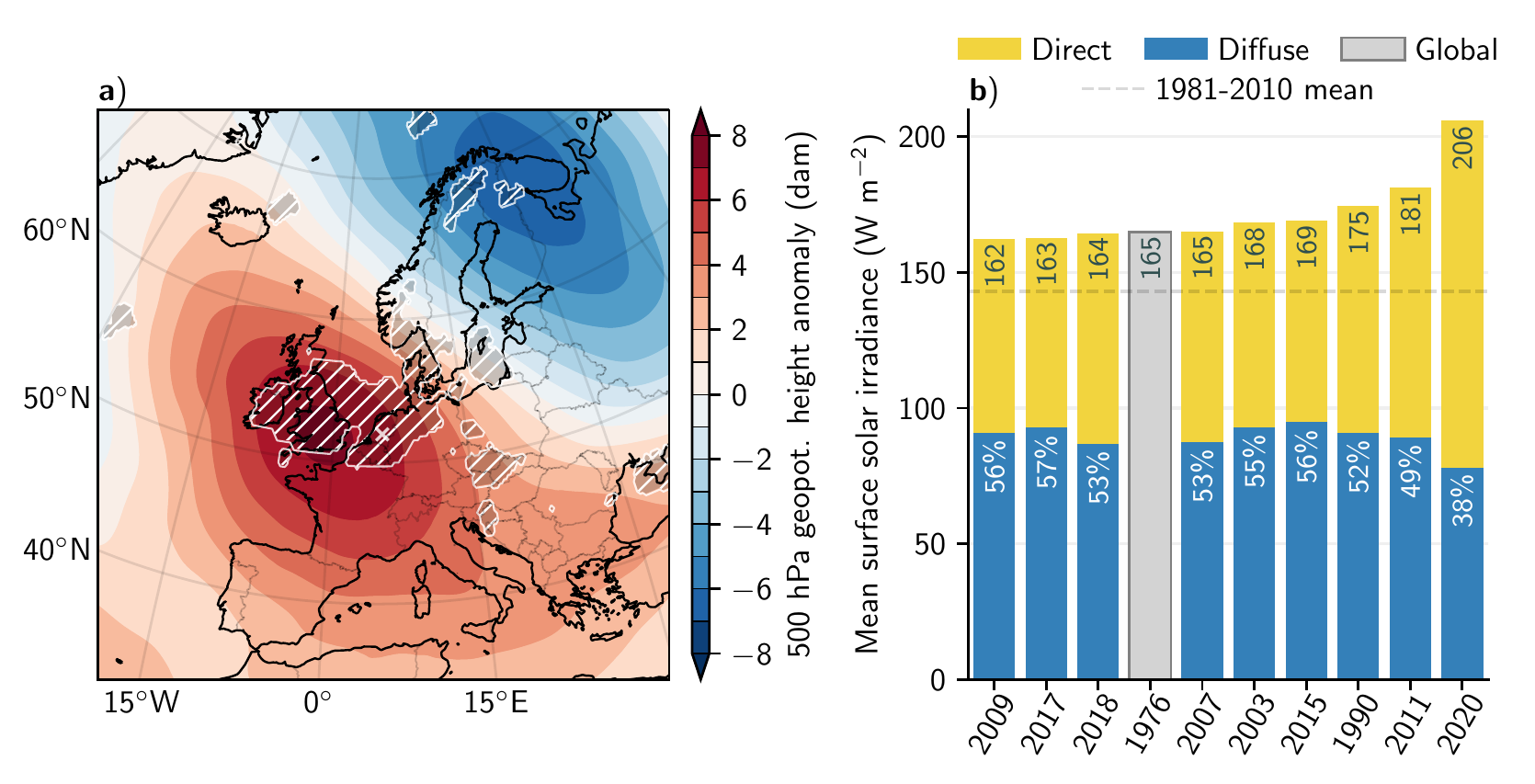}
\caption{\textbf{Surface irradiance in spring 2020 relative to earlier years.}
Figure shows \textbf{a)} 500 hPa geopotential height anomaly of 2020 spring (March, April, May) with respect to the 1981--2010 climatology, based on ERA5 reanalysis \cite{Hersbach2020}, hatched area indicate locations where the global irradiance (sum of direct and diffuse) in ERA5 exceeds the 1979--2019 maximum by more than 1\,\%, and \textbf{b)} top-10 years of daily mean integrated global horizontal irradiance (GHI) since 1928 for the Veenkampen station (white cross in panel a) from March 1 until May 31, partitioned into direct and diffuse. The percentages show the diffuse portion of GHI where such measurements are available. Veenkampen station was moved over a distance of 2 km in 2012, 1989 is missing from the record.}
\label{fig:radrecord}
\end{figure}

\section{Introduction}
A large part of Western Europe (Fig. \ref{fig:radrecord}a, hatched area) experienced exceptionally sunny and dry weather from March 23 to the end of May 2020.
Sunshine duration extremes were reported in the United Kingdom, Belgium, Germany, and The Netherlands \cite{UKMO2020, DWD2020, KNMI2020, KMI2020} paired with exceptionally deep blue skies \cite{Knap2020, Meirink2020}.
For instance, The Netherlands reported 805 h of sunshine, compared to 517 h normally and 62 h more than the previous record of 2011 \cite{KNMI2020}.
This resulted in a time-integrated surface solar irradiance for spring (March, April, May) that was the largest ever observed at the Veenkampen station (The Netherlands) since 1928 (Fig. \ref{fig:radrecord}b).
The daily mean irradiance sum of 206 W m$^{-2}$ exceeded the previous record of 2011 by 25 W m$^{-2}$.
The diffuse radiation reaching the surface was only 38\,\% of the total solar irradiance in the period, compared to 49--58\,\% in the other top-ten springs with high irradiance (Fig. \ref{fig:radrecord}).
Clouds reduced daily solar irradiance on average by 22\,\% in spring 2020 with respect to clear-sky conditions, $3 \sigma$ less than the 2004-2020 mean reduction of 36\,\%, as computed from the observations of solar irradiance from Veenkampen station and clear-sky surface solar irradiance taken from McClear \cite{Gschwind2019}.

These records all happened amid the first European wave of the COVID-19 pandemic \cite{JHU2020, ECDC2020}, during which many countries went into lockdown, leading to a global reduction in anthropogenic pollution \cite{Muhammad2020}.
Less traffic and industrial activity led to losses in NO$_x$, SO$_2$, and CO$_2$ emissions of tens of percents \cite{Bauwens2020, Forster2020, LeQur2020}, with consequent changes to the atmospheric composition \cite{Le2020, Kroll2020} and the radiation balance \cite{Forster2020}.
The large leap with which the irradiance records were broken made us hypothesize that the reduction in anthropogenic aerosols and contrails related to the COVID-19 lockdown are a secondary driving force behind the observed irradiance extremes next to the exceptionally cloud-free weather.
This study aims to quantify the individual contributions of weather and aerosols to the extreme irradiance of 2020.
With the already relatively clean air of Western Europe without the COVID-19 lockdown, we expect the contribution of weather to exceed that of aerosols.
Nonetheless, an exact quantification of the contribution will provide useful insight in the lockdown effects, as well as in the extremity of the weather and future surface irradiance extremes.

\begin{figure}[ht]
\centering
\includegraphics[width=1\textwidth]{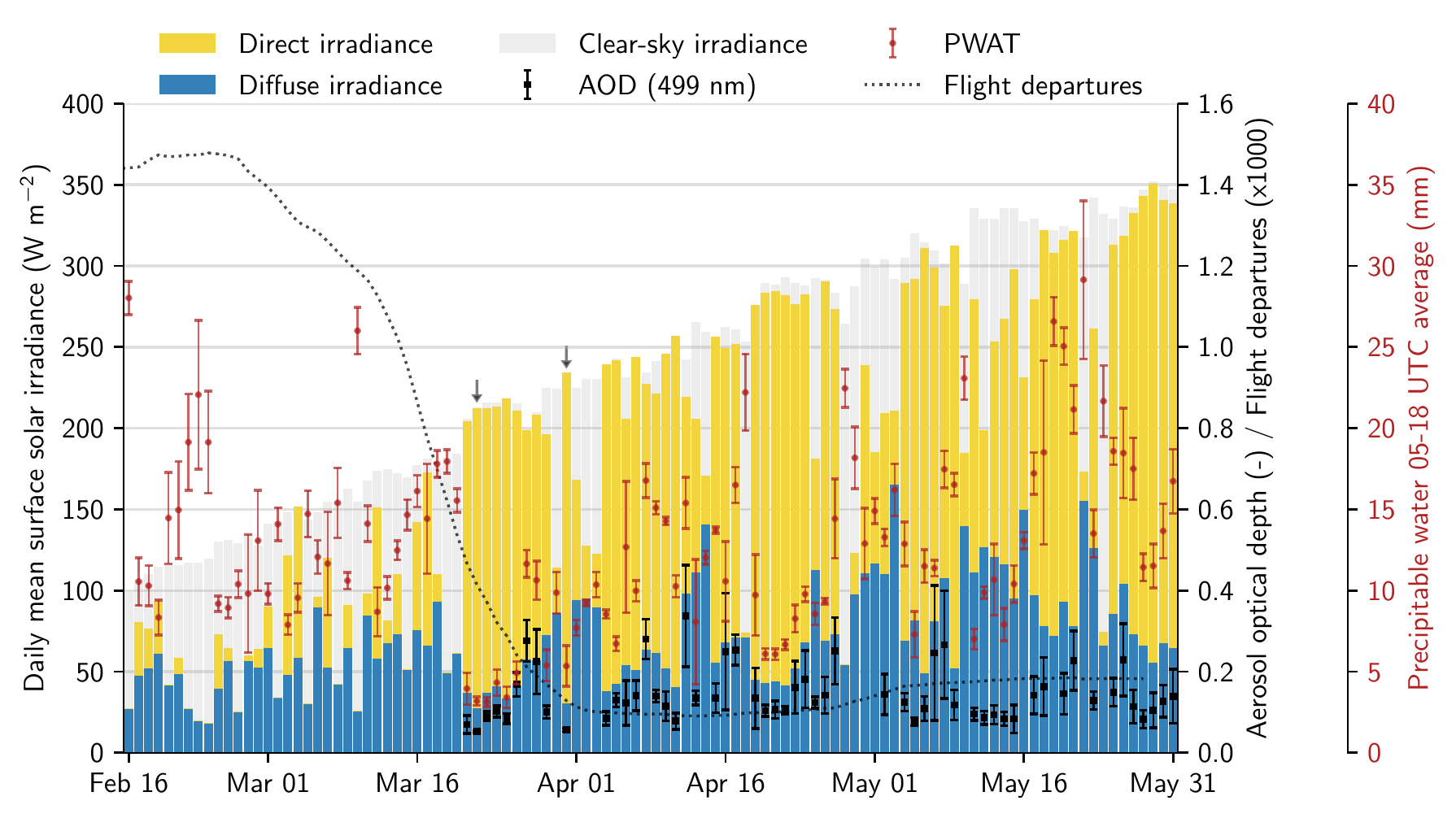}
\caption{\textbf{Time series of relevant variables in spring 2020 in Cabauw, NL}. These include time series of measured direct and diffuse irradiance (BSRN Cabauw, NL), clear-sky global horizontal irradiance (Copernicus Atm. Monitoring Service (CAMS) McClear dataset for Cabauw), 499 nm aerosol optical depth (AOD) at Cabauw measured using a precision filter radiometer and processed at PMOD/WRC \cite{Kazadzis2018}, daytime precipitable water at the grid point closest to Cabauw based on ERA5 reanalysis \cite{Hersbach2020} and the weekly moving average of flight departures at (major) Western Europe airports (OpenSkyNetwork COVID-19 dataset \cite{Schafer2014}). Error bars indicate daily variability ($\pm \sigma$). Arrows point to 22 and 31 March. Instrument maintenance and cloud-contamination prevented ground-based AOD observations before 21 March. Complementary AOD data combined with normalized irradiance and a cross-validation of the complementary AOD data are found in Supplementary Material Figs. \ref{fig:covid_overview_alternative} and \ref{fig:aod_cams_vs_cabauw}.}
\label{fig:radtimeseries}
\end{figure}

\section{Results}
\subsection{Time evolution of spring 2020}
To test our hypothesis that COVID-19 lockdowns have contributed to the irradiance extreme, we first analysed data \cite{Cabauw2020} from the Baseline Surface Radiation Network's (BSRN \cite{Driemel2018}) measurement station in Cabauw, The Netherlands.
This station is located in the center of the regions that reported sunshine duration records, and has already available observations of irradiance and aerosol optical depth (AOD) for spring 2020 (Fig. \ref{fig:radtimeseries}).
The onset of the prolonged time period of fair weather on March 21, as identified by the moment where the global irradiance starts approximating the clear-sky radiation, coincides with the strong drop in flight activity that marked the onset of the COVID-19 lockdown in many European countries (Fig. \ref{fig:radtimeseries}).
The fair weather is reflected by the large amounts of global irradiance, i.e., direct and diffuse irradiance taken together in the observations (Fig. \ref{fig:radtimeseries}), and the large contribution of direct solar irradiance therein.
Until May 31, there were only three overcast days.
The surface irradiance is gradually increasing over time towards the end of May, hence the sunny days later in the period weigh more heavily in the mean shown in Fig. \ref{fig:radrecord}.
To enable comparison among days without the yearly cycle, Fig. \ref{fig:covid_overview_alternative} in the Supplementary Material contains a normalized version of Fig. \ref{fig:radtimeseries}.

Especially the period of 21 to 31 March was remarkably cloud free, e.g., seen by the global irradiance equal to the clear-sky radiation, and values for diffuse irradiance are the smallest in the period.
These days recorded the lowest AOD of the entire period and the lowest precipitable water in the atmosphere (Fig. \ref{fig:radtimeseries}), underlining the cleanliness and dryness of the air.
Radiosonde observations of De Bilt showed strikingly low amounts of precipitable water (Supplementary Material, Fig. \ref{fig:ehdb_sounding}).
Based on ERA5 Reanalysis at a similar location, March 22 to 26 had on average 4.0 $\pm$ 1.2\ kg\ m$^{-2}$ precipitable water, far below the 1981--2019 mean of 11.5 $\pm$ 4.2\ kg\ m$^{-2}$ for the same period.
Later, in May, multiple days had a very low AOD (below 0.1), including days with partial cloudiness, e.g., May 11 to 15.

\begin{figure}[ht]
\centering
\includegraphics[width=1\textwidth]{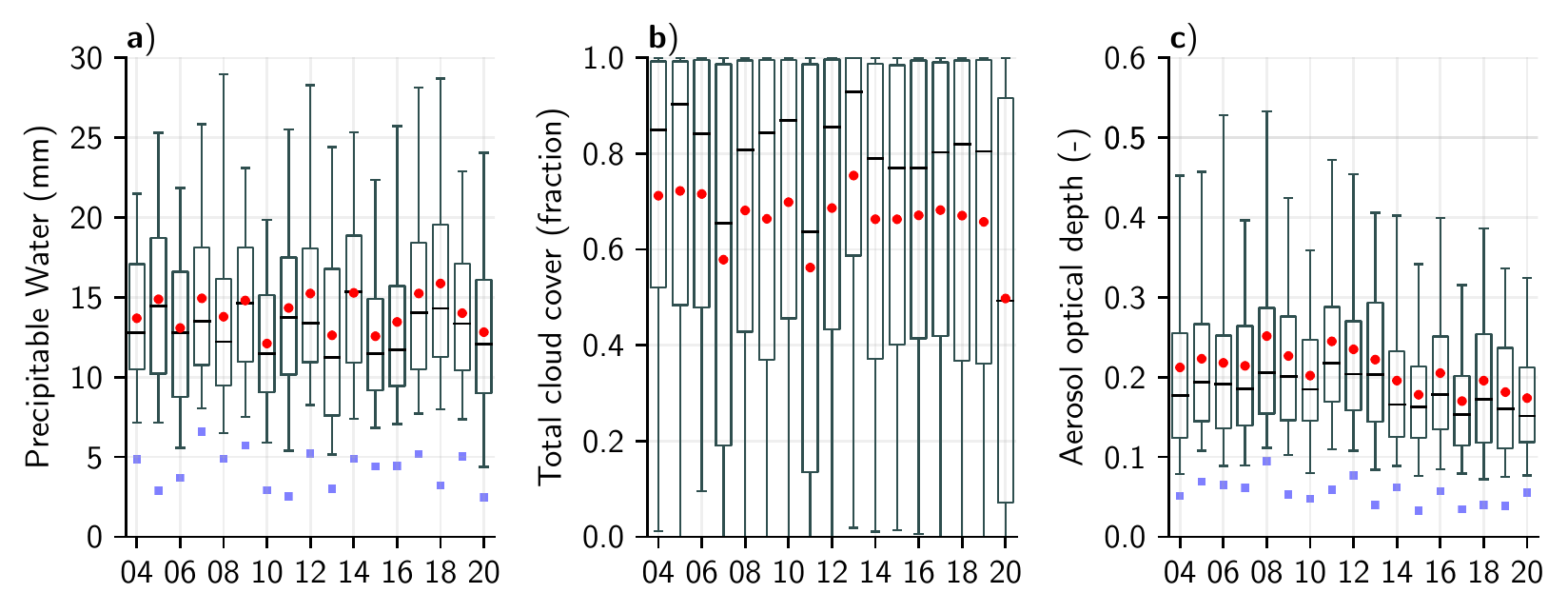}
\caption{\textbf{Box plots of values during spring of variables affecting surface solar irradiance.} Figure shows \textbf{a)} ERA5 precipitable water, \textbf{b)} ERA5 total cloud cover, and \textbf{c)} CAMS aerosol optical depth at 550 nm. Box plots are based on hourly values for March, April and May, between 5 and 18 UTC (approximate mean day time) at the location or grid point closest to Cabauw, NL. Red circles indicate mean values, blue squares are minima, and whiskers represent 5th and 95th percentiles. Data spans years 2004 to 2020.}
\label{fig:anomaly}
\end{figure}

\subsection{Anomalies in weather patterns, aerosols, and contrails}
In order to assess the potential impact of the COVID-19 lockdown on the irradiance extremes, and to evaluate extremes in weather versus human activity, we discuss the anomalies in atmospheric circulation (Fig. \ref{fig:radrecord}a), in precipitable water, cloud cover, and AOD (Fig. \ref{fig:anomaly}), and in contrail formation with respect to their climatology.

Spring 2020 had recurring weather patterns favorable for sunshine, with persistent north- to easterly flow over Western Europe or weak winds in the centre of high pressure systems.
These conditions are reflected by a positive anomaly in the 500-hPa geopotential height in the lockdown period centered over the south of the United Kingdom and a negative anomaly in northern Scandinavia and Russia (Fig. \ref{fig:radrecord}a).
This pattern is typical for atmospheric blocking conditions \cite{Folland2009}, which are often drivers of heatwaves in summer and cold spells in winter \cite{Pfahl2012}.
In absence of temperature extremes, springtime blocking conditions attract relatively less interest \cite{Woollings2018}, despite regular occurrence \cite{Brunner2017}.
They are, as this study shows, a contributor to surface irradiance extremes due to their related cloud-free skies, caused by its dry and sinking air masses.
To further elaborate the weather patterns, the German objective weather type classification \cite{Bissolli2001}, which contains The Netherlands within its domain, showed that 2020 had approximately ten more spring days in dry and anticyclonic regimes compared to its 1980-2019 mean (Supplementary Material Fig. \ref{fig:dwd_class}), confirming favorable synoptic conditions for sunshine.

Also the precipitable water and total cloud cover (Fig. \ref{fig:anomaly}a and Fig. \ref{fig:anomaly}b) display the exceptional weather.
The box plot of hourly values of precipitable water (Fig. \ref{fig:anomaly}a), which measures the vertical integral of water in the atmosphere, shows that 2020 was among the drier years in the record (2004-2020) with the lowest 5th-percentile and minimum value corresponding to the very dry period starting March 21 (Fig. \ref{fig:radtimeseries}).
The exceptional conditions of 2020 are most prominently reflected in the box plot of hourly values of total cloud cover (Fig. \ref{fig:anomaly}b), with the 95th-percentile, mean, median, and 5th-percentile being by far the lowest on record. The mean cloud cover of 0.5 is more than 0.16 less than the mean over 2004-2020 (0.66) and also 0.05 lower than 2011, the year of the previous record.

Within cloud-free conditions, irradiance increases with decreasing AOD.
We documented many days in spring 2020 with exceptionally low humidity and a low AOD (Fig. \ref{fig:radtimeseries}).
To appreciate the AOD observations, we have to acknowledge the challenge in separating aerosols from homogeneous, optically thin cirrus in observations using sun photometers \cite{Chew2011} or satellite products \cite{Kaufman2005}, which can result in a positively biased AOD.
To avoid this difficulty and to assess a longer statistic of AOD, we use here the hourly AOD values from the CAMS aerosol product (Fig. \ref{fig:anomaly}c, Supplementary Material Fig. \ref{fig:covid_overview_alternative}), which compares well against ground-based observations over Europe \cite{Gueymard2020} (Supplementary Material Fig. \ref{fig:aod_cams_vs_cabauw}).
Our analysis highlights that spring 2020 had the lowest median in hourly AOD since 2004.
Spring 2020 was also among the springs with the lowest mean AOD, with only 2015 and 2017 being lower, but its minimum and 5th percentile values did not stand out from the statistics of hourly AODs.
Therefore, the reduction in anthropogenic aerosol pollution due to the lockdowns did not lead to new extremely low hourly values of AOD, but rather to frequent hours with low AODs.

A reduction in contrail-cirrus due to the drop in flight activity (Fig. \ref{fig:radtimeseries}) is another pathway for the lockdowns to enhance surface irradiance.
This is particularly true in Western Europe, which is a hot spot for contrail-cirrus, owing to a combination of high aviation activity, and suitable meteorological conditions.
To obtain a rough estimate of the effect, we compared spring 2020 to 2011 and 2015, which are both among the top-five years in terms of surface irradiance (Fig. \ref{fig:radrecord}b), while having contrasting AODs.
Year 2011 had the highest median AOD in recent years, whereas 2015 had an AOD statistic comparable to 2020 (Fig. \ref{fig:anomaly}).
The meteorological conditions for persistent contrail formation at 250 hPa, close to the typical flight level of 230 hPa \cite{Burkhardt2011, Lee2009, Jensen1998}, are slightly less favourable (see Supplementary Table \ref{tab:modis} and Fig. \ref{fig:contrail_conditions}), thus we expect to observe less contrail-cirrus in 2020.
Manual inspection of cirrus and contrail occurrence (see Methods and Fig. \ref{fig:contrailcirrus} for an example) in NASA Worldview \cite{NASA2020} imagery for The Netherlands gives results consistent with our expectation.
The images showed that 2011 and 2015 had about twice as much cirrus, but with 50\,\% more contrail contamination, compared to 2020.
Given that contrail-cirrus has a net shortwave radiative forcing in the order of -1 W m$^{-2}$ over Western Europe \cite{Stuber2006}, but can enhance diffuse irradiance with tens of percents \cite{Feister2005, Gueymard2012, Weihs2015}, we conjecture that the low presence of contrails contributed to the extremely low diffuse fraction that was observed in 2020.
To illustrate this further, we selected three pairs of consecutive days with a clear sky on the first day and cirrus clouds on the second.
Clear-sky days March 23, April 22, and May 6 had diffuse fractions of respectively 17, 14, and 16\,\%, whereas days with cirrus March 24, April 23, and May 7 showed diffuse fractions of 19, 24, and 27\,\%.
In each of the three pairs, the irradiance relative to clear-sky dropped at most a percent on the day with cirrus, indicating that cirrus clouds mainly caused downward scattering.
This provides further evidence that, despite its large influence on diffuse irradiance, the reduced aviation is likely to play a minor role in the global surface irradiance extreme.

\begin{figure}[ht]
\centering
\includegraphics[width=\textwidth]{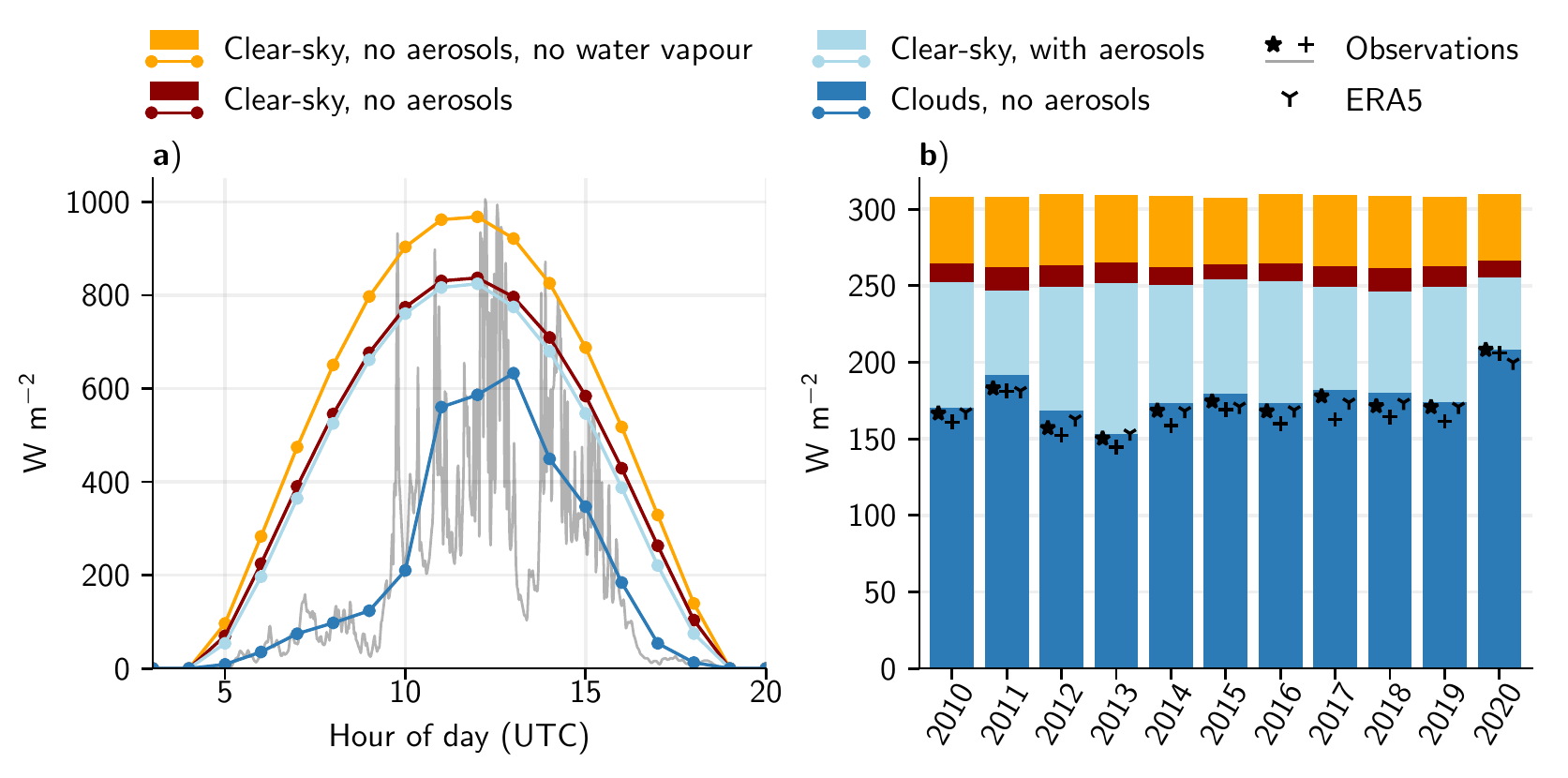}
\caption{\textbf{Modelled surface irradiance with a radiative transfer model}. Irradiance is shown for \textbf{a)} an illustrative case study of 29 April, 2020 compared against Cabauw observations, and \textbf{b)} averaged over the months March, April, and May compared against Cabauw ($\star$) and Veenkampen ($+$) observations and against the surface irradiance of ERA5. See Table \ref{tab:subl_rad_experiment} in Supplementary Table \ref{tab:subl_rad_experiment} for exact values corresponding to bars.}
\label{fig:radmodel}
\end{figure}

\subsection{Radiative transfer modelling for interpretation of extremes}
We present here estimations of i) the relative importance of different contributing factors to the extreme in surface irradiance, and ii) the anomalies in cloud radiative forcing and the direct aerosol effects.
To this end, we used a contemporary radiative transfer model \cite{Pincus2019} to first reproduce the observed surface irradiance, and subsequently repeat the calculation without individual components to assess their quantitative contribution to the surface irradiance.
We used the McClear clear-sky radiation product to infer the aerosol contribution \cite{Gschwind2019} (see Supplementary Fig. \ref{fig:aod_validation} for a validation of the direct aerosol effect as a function of AOD).
Further details are given in the Methods section.
The combination of the experiments with the radiative transfer model and the clear-sky data provides surface irradiance under four conditions, compared against the observations in Fig. \ref{fig:radmodel}: i) experiment \texttt{dark blue} resembles the reality, but without aerosols in the atmosphere, ii) experiment \texttt{red} additionally removes the clouds, iii) experiment \texttt{orange} additionally removes the water vapour, and iv) experiment \texttt{light blue} is the clear-sky product, thus without clouds but with aerosols.

We first show the modelled surface solar irradiance for a single day on 29 April in Fig. \ref{fig:radmodel}a for illustrating the transient behaviour of the radiative transfer model results.
The experiment \texttt{dark blue} with clouds, but without aerosols closely follows the observed slowly increasing irradiance due to vanishing clouds.
It confirms the ability of the model to reproduce the time series of surface irradiance (see Supplementary Fig. \ref{fig:rrtmgp_validation} for validation).
At noon, removal of clouds (\texttt{red}) increases the irradiance by 250.7 W m$^{-2}$, and removing water vapour (\texttt{orange}) increases irradiance by a further 130.8  W m$^{-2}$, whereas the presence of aerosols (\texttt{light blue}) lowers the irradiance only by 12.5 W m$^{-2}$.
Both the removal of water vapour and clouds have a larger effect on the irradiance than removing aerosols, due to the typically larger optical depth of clouds than of aerosols.

We expand our analysis to the entire spring period for each of the past 10 years in Fig. \ref{fig:radmodel}b.
Here, the top of each bar segment indicates the global surface irradiance for the situation indicated by its color.
If clouds, aerosols, and water vapour are removed, all years have a bar of approximately equal depth, indicating that our experiment captures the essence and that leap years and year-to-year variability in atmospheric pressure, ozone, and temperature are of minor importance to the irradiance extremes in 2020.
The variation over the years is comparable between the two measurement stations and is closely following the radiative transfer computations with clouds (\texttt{dark blue}), but without aerosols.
The observations are consistently lower than the model simulation, consistent with the here removed aerosols.
Furthermore, the Cabauw station has a consistently higher irradiance than Veenkampen, due to its closer proximity to the coast, where clouds are less common \cite{KNMIsunshine2020}.

We quantify the cloud radiative effect at the surface as the difference between the experiment with clouds and the clear-sky experiment without aerosols (\texttt{dark blue} minus \texttt{red}).
The irradiance increase due to the reduction in clouds is +30.7 W m$^{-2}$ in spring 2020 with respect to the 2010--2019 mean cloud radiative effect of -88.6 W m$^{-2}$.
Similarly, we quantify the aerosol effect from the difference between the clear-sky experiment without aerosols and the McClear data (\texttt{light blue} minus \texttt{red}).
This indicates an increase in irradiance of +2.3 W m$^{-2}$, with respect to the 2010--2019 mean aerosol effect of -13.0 W m$^{-2}$.
The water vapour effect is quantified as the clear-sky experiment minus the dry experiment (\texttt{red} minus \texttt{orange}) and gives an enhancement of only +1.5 W m$^{-2}$ with respect to the 2010--2019 mean water vapour effect of -45.4 W m$^{-2}$.
The vapour enhancement is only the contribution to the optical properties of the clear-sky radiation, and that the most important signal of the atmospheric moisture anomaly of 2020 is tightly linked with the low magnitude of the cloud cover.
The low humidity also affects the aerosol optical depth in the sense that less water vapour is available for condensation on the aerosol surface keeping the aerosol optical depth smaller than in moist conditions.
The quantification of the three effects highlights the relative importance of variations in cloud cover over the years in explaining the surface irradiance.
It emphasizes that the sunny weather played the most important role in setting the 2020 record in surface irradiance, while the reduced emission of anthropogenic aerosols is of smaller importance to the extent that even without the reduction the irradiance record had occurred.

\section{Discussion}
During the exceptionally sunny spring in Western Europe amid the COVID-19 pandemic in 2020, The Netherlands received the most solar radiation at the surface since the start of the measurements in 1928 and never experienced so little scattering of light (Fig. \ref{fig:radrecord}b).
The particularly dry atmosphere (Fig. \ref{fig:radtimeseries}) and weather patterns favouring sunny weather (Fig. \ref{fig:radrecord}a) led to fewer clouds than in previous years.
Based on radiative transfer calculations, we estimated the relative contributions of clouds, water vapour, and aerosols and argue that the former is the dominant contributor to the new irradiance extreme (Figs. \ref{fig:anomaly} and \ref{fig:radmodel}), while the impact of COVID-19 measures via aerosols is an order of magnitude less.

With all but two of the top-10 spring irradiance years since 1928 observed in the two most recent decades (Fig. \ref{fig:radrecord}), it has become clear that conditions are now more favorable for sunshine than in the past.
This is partially explained by a well-documented downward trend in aerosol concentrations over Western Europe, leading to a large-scale upward trend in surface solar irradiance, often referred to as brightening \cite{Wild2004, Wild2005, Wild2012}.
Our data fits into that picture as, measured by the median, spring 2020 was the cleanest on record in The Netherlands since 2004 (Fig. \ref{fig:anomaly}b) and the data shows an ongoing downward AOD trend (-3.5\,$\cdot 10^{-3}$ y$^{-1}$, as derived from Fig. \ref{fig:anomaly}b).
With exceptional weather conditions emerging as the main contributor to the 2020 extreme, the question arises if weather patterns are also showing a trend that is more favorable for sunshine.
There is a clear link between weather patterns and surface irradiance, as has been shown for Northern Europe \cite{Parding2016}.
At the same time, there is an extensive debate on the existence of trends in weather patterns in Europe in recent decades, mainly due to their strong sensitivity to the chosen classification method and domain location and size \cite{Kucerova2017}.
To illustrate, an increase in spring blocking highs over Europe and a drop in cyclonic activity in Scandinavia has been reported \cite{Philipp2007}, in line with more general claims that anthropogenic influence on the jet stream makes weather more persistent \cite{Francis2012, Coumou2018}.
At the same time, recent studies show an absence of trends in the frequency and the persistence of blocking events in spring \cite{Brunner2017, Huguenin2020}.
With the above-mentioned uncertainty in trends, we cannot confirm nor rule out the influence of changing weather patterns on the frequency of cloud-free and exceptionally dry skies.
The still uncertain response of clouds and circulation to warming \cite{Woollings2018, Bony2015} prevents a conclusion on their future contribution to the frequency and strength of new seasonal extremes in irradiance.

Further effects of the aerosol removal are possible, including rapid adjustments of clouds to associated temperature changes and aerosol effects on cloud microphysical processes, but these can not be quantitatively assessed with the model used here.
Furthermore, given the worldwide impact of COVID-19 lockdown on emissions of anthropogenic aerosols and greenhouse gases \cite{Forster2020}, there exists a possibility of a response of the global circulation to the perturbed radiation balance.
With the inherent variability of the weather, proving such a response is challenging based on a single spring, yet studies with global models have shown interactions between aerosols, clouds, and circulation based on longer time scales \cite{Liepert2004, Liu2019, Wilcox2019, Fiedler2020}.
To conclude, if we account for additional aerosol effects, the impact of the reduced anthropogenic aerosols due to the lockdown is potentially larger than our estimate of the instantaneous clear-sky radiative effects of the aerosol removal.

With the prospect of anthropogenic aerosol emissions in Europe to stay small or even decrease further in the future \cite{Fiedler2019}, weather will be the most relevant factor in establishing new spring irradiance records.
At the moment, however, many regions of the world are more strongly polluted than Western Europe, and larger regional effects of aerosol reductions on irradiance from COVID-19 lockdowns are already being documented \cite{Peters2020}.
With the pandemic still going on, more data to test our expectation will become available in the near future.

\section*{Author contributions}
CvH, WM, and MV contributed equally.
CvH, WM, and MV designed the study.
WM performed the data analysis of the observations.
MV performed the radiative transfer model experiments.
CvH, WM, MV, IB and SF interpreted the results.
CvH, WM, MV, and SF wrote the manuscript.
BH, WK, SK, and NK provided observational data and expertise thereof.
All authors read the final manuscript and provided feedback.

\section*{Acknowledgments}
CvH, WM, MV, and BH acknowledge funding from the Dutch Research Council (NWO) (grant: VI.Vidi.192.068). SF acknowledges the funding of the Hans-Ertel-Centre for Weather Research by the German Federal Ministry for Transportation and Digital Infrastructure, (grant: BMVI/DWD 4818DWDP5A).

\section{Methods}
\subsection{Cirrus and contrail-contamination estimation}
Effects of aviation are estimated by comparing two top-5 irradiance years with relatively high and low AOD with 2020, hence the choice of 2011 and 2015.
Environmental conditions favourable for persistent contrails, namely low air temperature at flight level \cite{Jensen1998}, and often occurring supersaturation with respect to ice \cite{Burkhardt2011, Lee2009}, are quantified using ERA5 reanalysis for a domain covering approximately The Netherlands at 250 hPa.

Cirrus occurrence and whether it is contaminated by contrails is manually counted by looking at high resolution satellite images from Terra and Aqua MODIS, available on the NASA Worldview website \cite{NASA2020}.
The inspected area covers The Netherlands and closely neighbouring regions (Belgium, western Germany and part of the North Sea).
This area is part of the irradiance extreme coverage (Fig. \ref{fig:anomaly} and helps offset the fact only two images close to noon per day are available.
Only cirrus and contrails optically thick enough to be detectable by eye can be counted, anything that is too thin to detect is assumed to have only a very small impact on irradiance.
Contrail-contamination is counted when there is cirrus with five or more linear (typically overlapping) or unnatural looking (dispersed) condensation trails present.
See Supplementary Fig. \ref{fig:contrailcirrus} for a clear example.

\subsection{Radiative transfer modelling}
We used the Radiative Transfer for Energetics and RRTM for General circulation model applications—Parallel (RTE+RRTMGP) \cite{Pincus2019} to reproduce the surface irradiance observations at Cabauw, The Netherlands for spring 2020 (March, April, May) in order to construct Fig. \ref{fig:radmodel}.
The computation requires hourly atmospheric profiles of pressure, temperature, water vapour, liquid water, ice, cloud cover, cloud liquid water, cloud ice, and ozone at a $0.25 \times 0.25^\circ$ grid resolution taken from the ERA5 reanalysis \cite{Hersbach2020}.
We used the data on 37 pressure levels instead of the 137 native model levels, but the vertical integrals are approximately conserved.
We assume i) clouds to be horizontally homogeneous within one grid cell,
ii) that adjacent cloud layers have overlap, iii) that the spatial correlation between two clouds layers decreases exponentially (using a decorrelation length of 2 km) with increasing vertical distance between the layers, and iv) that separated cloud layers have random overlap.
To obtain a statistical distribution of the cloud fields, we sampled 100 vertical profiles, calculated radiative fluxes for each profile and subsequently averaged the surface irradiance.
To infer the effect of aerosols, we used the surface irradiance product of the Copernicus Atmosphere Monitoring service (CAMS) McClear Clear-Sky Irradiating service \cite{Gschwind2019}.

\section{Data availability}
All data used in this manuscript is either available from public sources or included with this manuscript. Public sources are cited in the manuscript. Other data is available via \url{https://doi.org/10.5281/zenodo.4455892}.

\bibliographystyle{naturemag}

\setcounter{figure}{0}
\makeatletter
\renewcommand{\thefigure}{S\@arabic\c@figure}
\renewcommand{\thetable}{S\@arabic\c@table}
\makeatother

\title{\textbf{Supplementary material:} Record high solar irradiance in Western Europe during first COVID-19 lockdown largely due to unusual weather}

\maketitle
The supplementary material contains a series of figures and tables with additional material to complement and validate the analyses and radiative transfer model computations that are discussed in the main text. All items are referenced from the main text.
\clearpage

\begin{figure}[ht]
\centering
\includegraphics[width=1\textwidth]{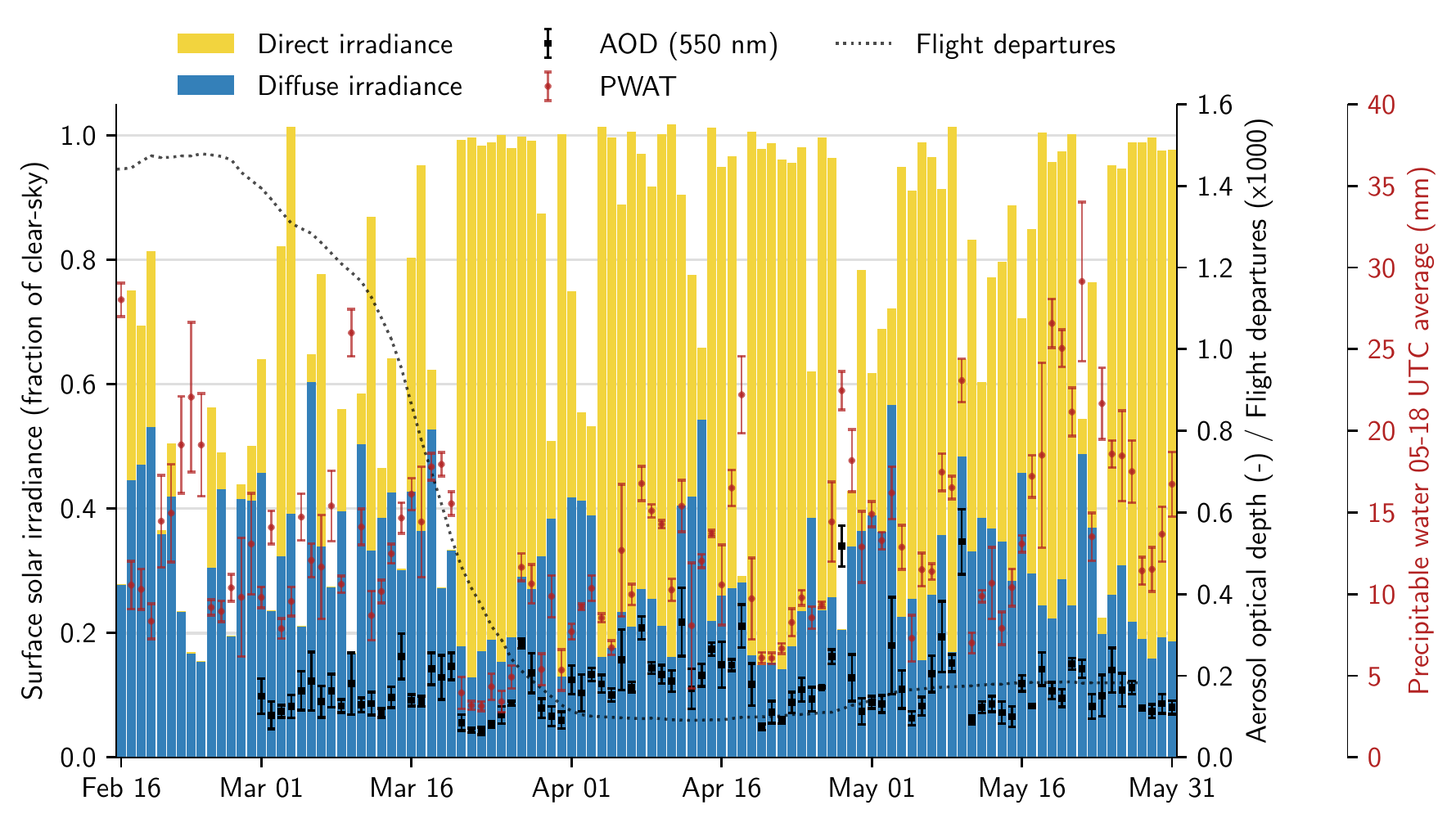}
\caption{Alternative version of Figure \ref{fig:radtimeseries}, where measured AOD has been replaced by CAMS McClear AOD (7 to 17 UTC mean + standard deviation) and solar irradiance measurements have been normalized by the CAMS McClear clear-sky global horizontal irradiance.}
\label{fig:covid_overview_alternative}
\end{figure}

\begin{figure}[ht]
\centering
\includegraphics[width=1\textwidth]{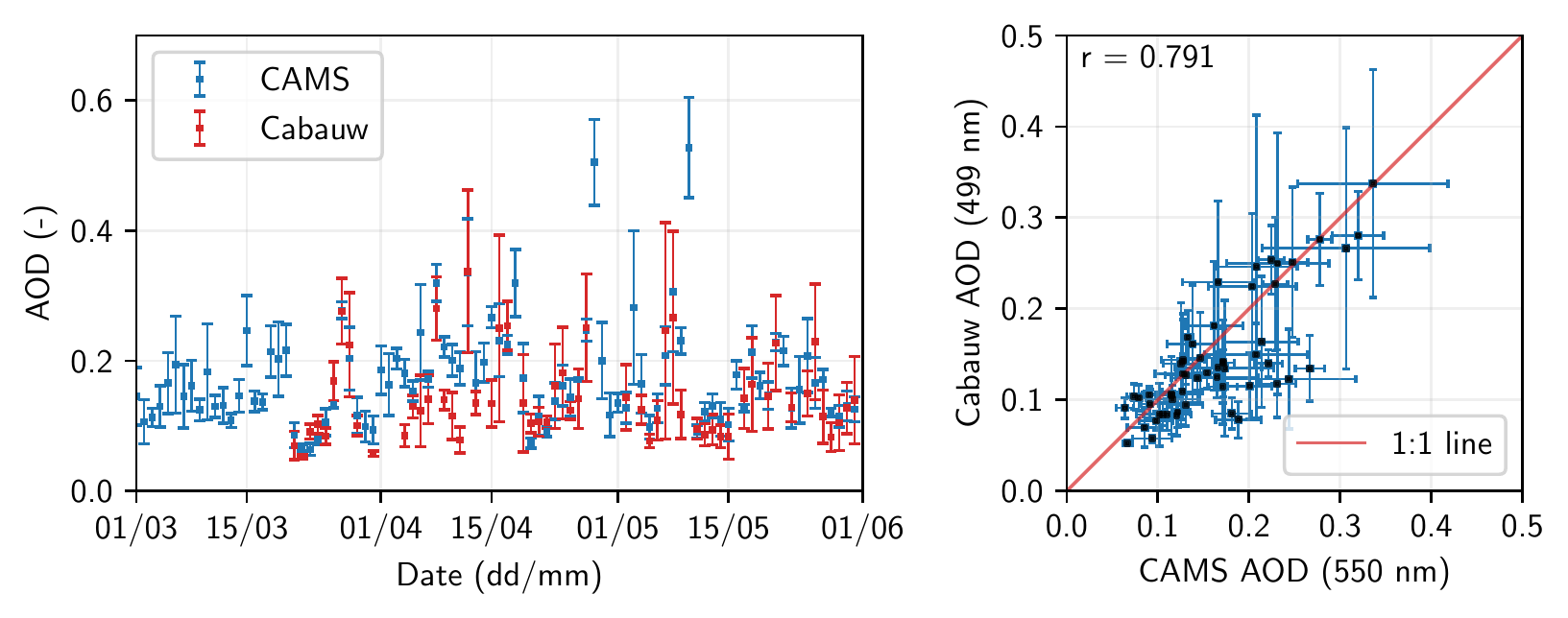}
\caption{Comparison of measured aerosol optical depth (AOD) of Cabauw and modelled AOD from CAMS for the same location during Spring 2020. Note Cabauw has missing data on cloudy days and for the first weeks of March. Errorbars indicate daily (daytime) variability in the data for each dataset. Correlation coefficient is 0.791 based on daily means for all days where both data sources have data available. Note Cabauw and CAMS AOD are at different wavelengths (499 vs. 550 nm), and daytime averages of Cabauw are based on a variable amount of samples due to irregular data availability (cloud contamination).}
\label{fig:aod_cams_vs_cabauw}
\end{figure}

\begin{figure}[ht]
\centering
\includegraphics[width=0.6\textwidth]{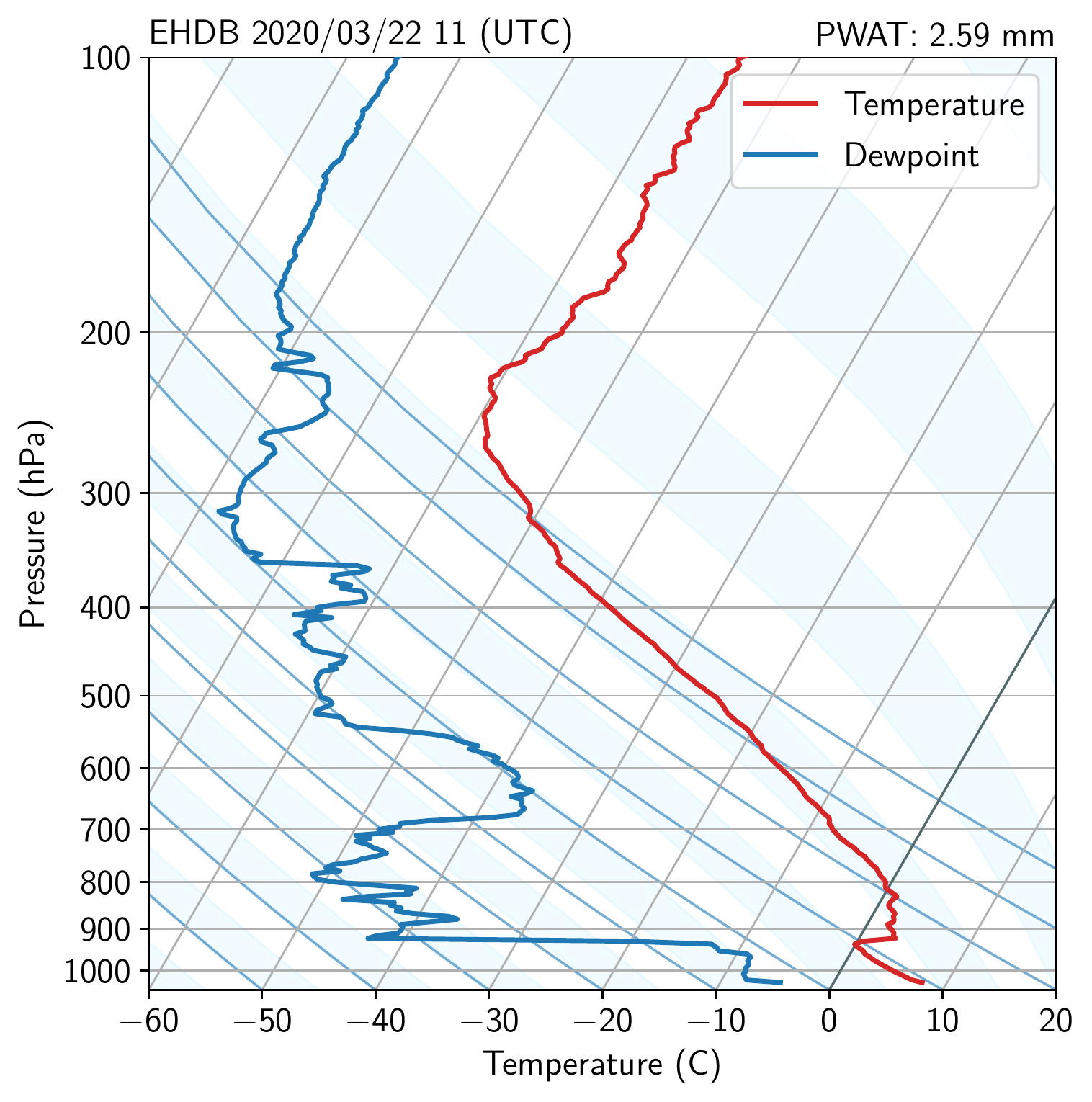}
\caption{Measured atmospheric profile on 22 March, 2020 at 11 UTC, De Bilt, The Netherlands. Vertically integrated moisture amounts to 2.59 mm. Data provided by the KNMI.}
\label{fig:ehdb_sounding}
\end{figure}

\begin{figure}[ht]
\centering
\includegraphics[width=1\textwidth]{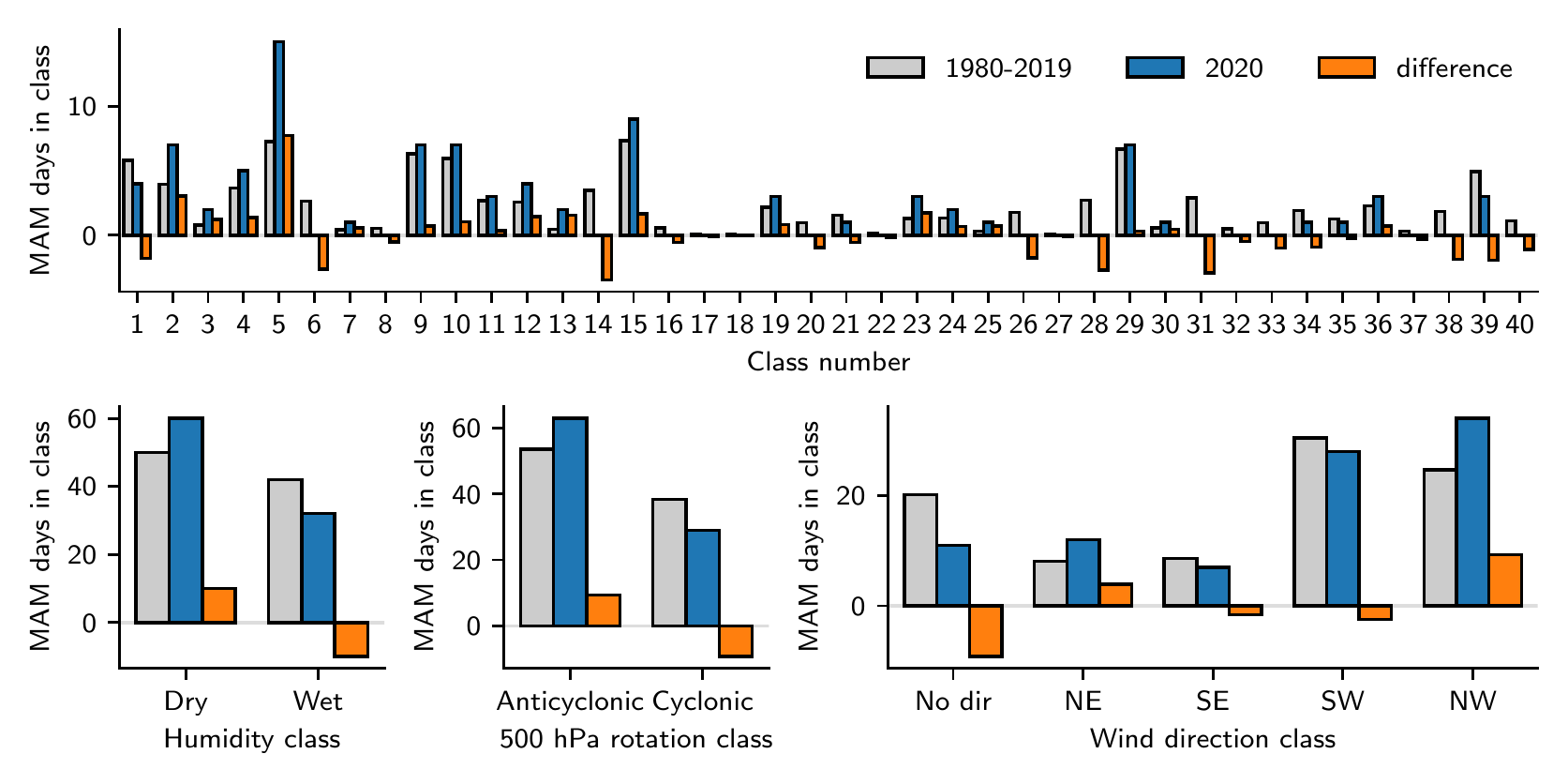}
\caption{Weather type classification of spring 2020 using the 40 groups of the German objective classification system \cite{Bissolli2001} compared against the 1980-2019 climatology of March, April, and May (MAM). Bottom row contains aggregated data from top row based on humidity, rotation at the 500 hPa level, or wind direction. See \url{https://www.dwd.de/EN/ourservices/wetterlagenklassifikation/kennzahlen_kennungen.html} for a description of the classes.}
\label{fig:dwd_class}
\end{figure}

\begin{table}[ht]
\centering
\caption{Percentage of days in March, April and May with visible cirrus and the percentage those days which are visibly contaminated with contrails. Based on Terra and Aqua MODIS imagery on NASA Worldview \cite{NASA2020}.}\label{tab:modis}
\begin{tabular}{l|ccc}
                    & 2011  & 2015  & 2020 \\
    \hline
    Cirrus          & 51 & 60 & 30 \\
    Contrail contaminated & 62 & 59 & 42 \\

\hline
\end{tabular}
\end{table}

\begin{figure}[ht]
\centering
\includegraphics[width=0.7\textwidth]{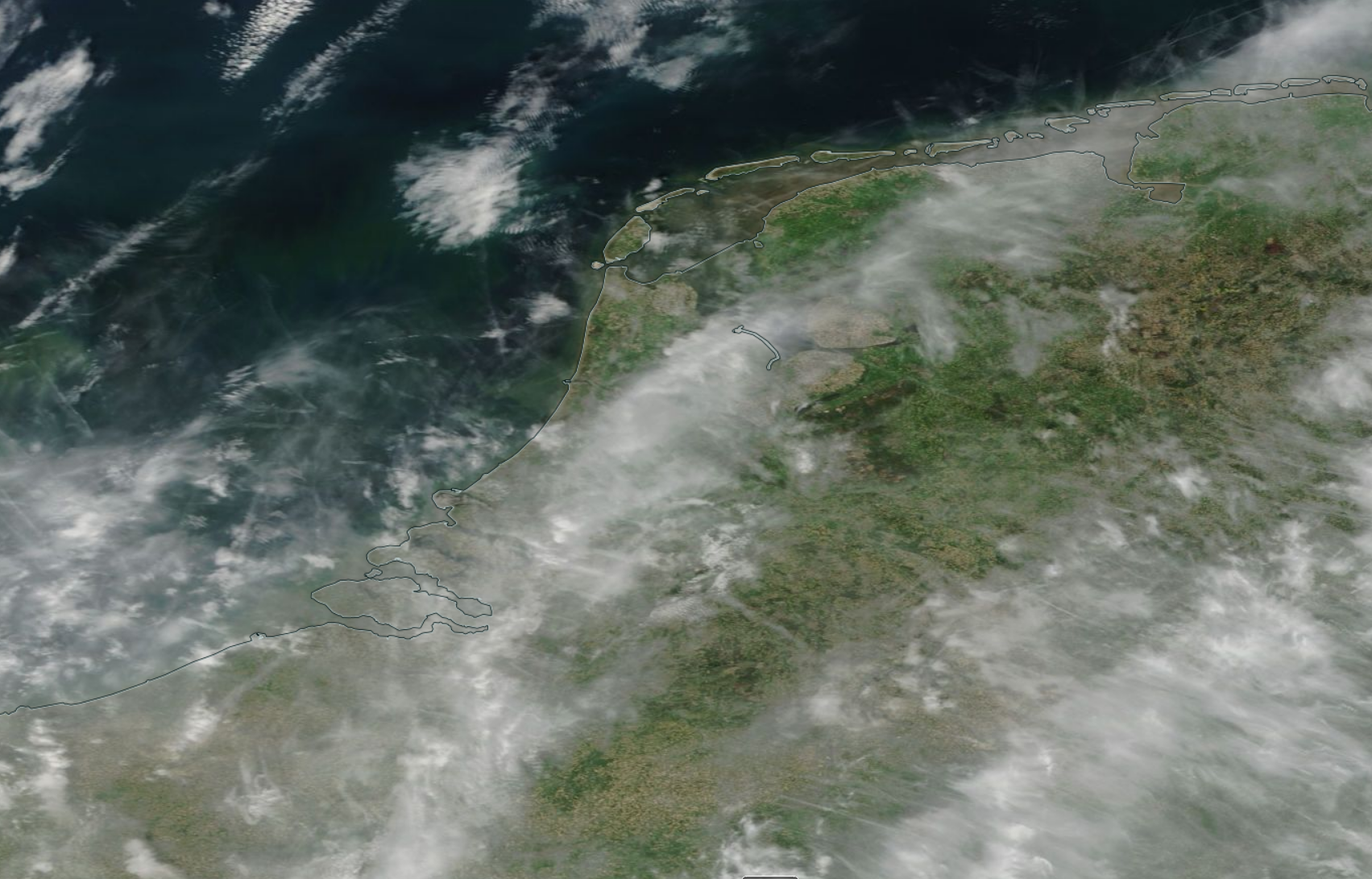}
\caption{Example from NASA Worldview of cirrus that contains (and is therefore enhanced in coverage and optical thickness) contrails, from narrow and straight ones to older and more dispersed. The date of this image is 11 May 2015.}
\label{fig:contrailcirrus}
\end{figure}

\begin{figure}[ht]
\centering
\includegraphics[width=\textwidth]{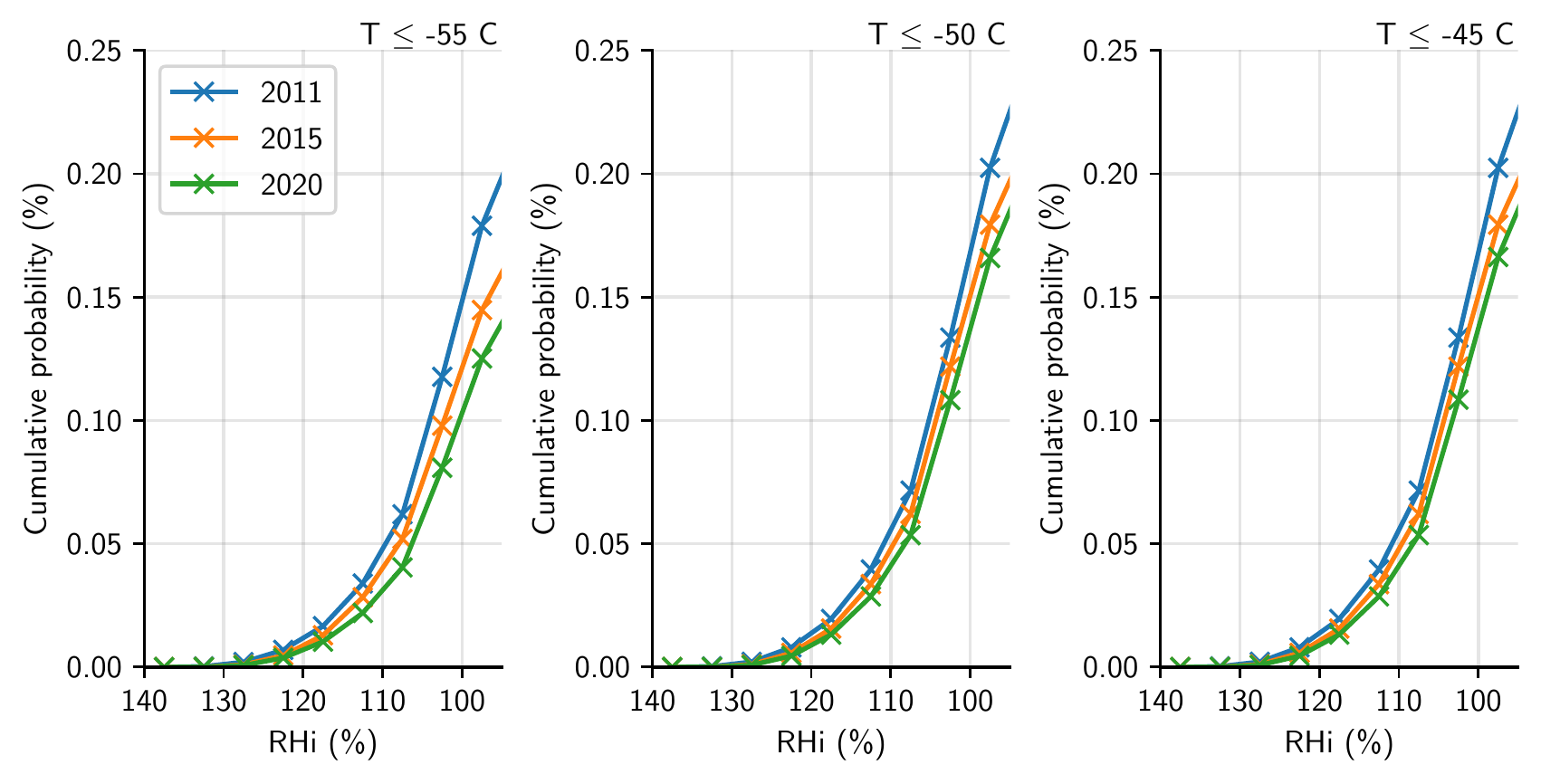}
\caption{Cumulative probability of relative humidity over ice (RHi) based on ERA5 Reanalysis data for three different temperature thresholds. Data is taken at 250 hPa from 50 to 55 degrees latitude and 3 to 7 degrees longitude and compares years 2020 to 2011 and 2015.}
\label{fig:contrail_conditions}
\end{figure}

\begin{figure}[ht]
\centering
\includegraphics[width=0.9\textwidth]{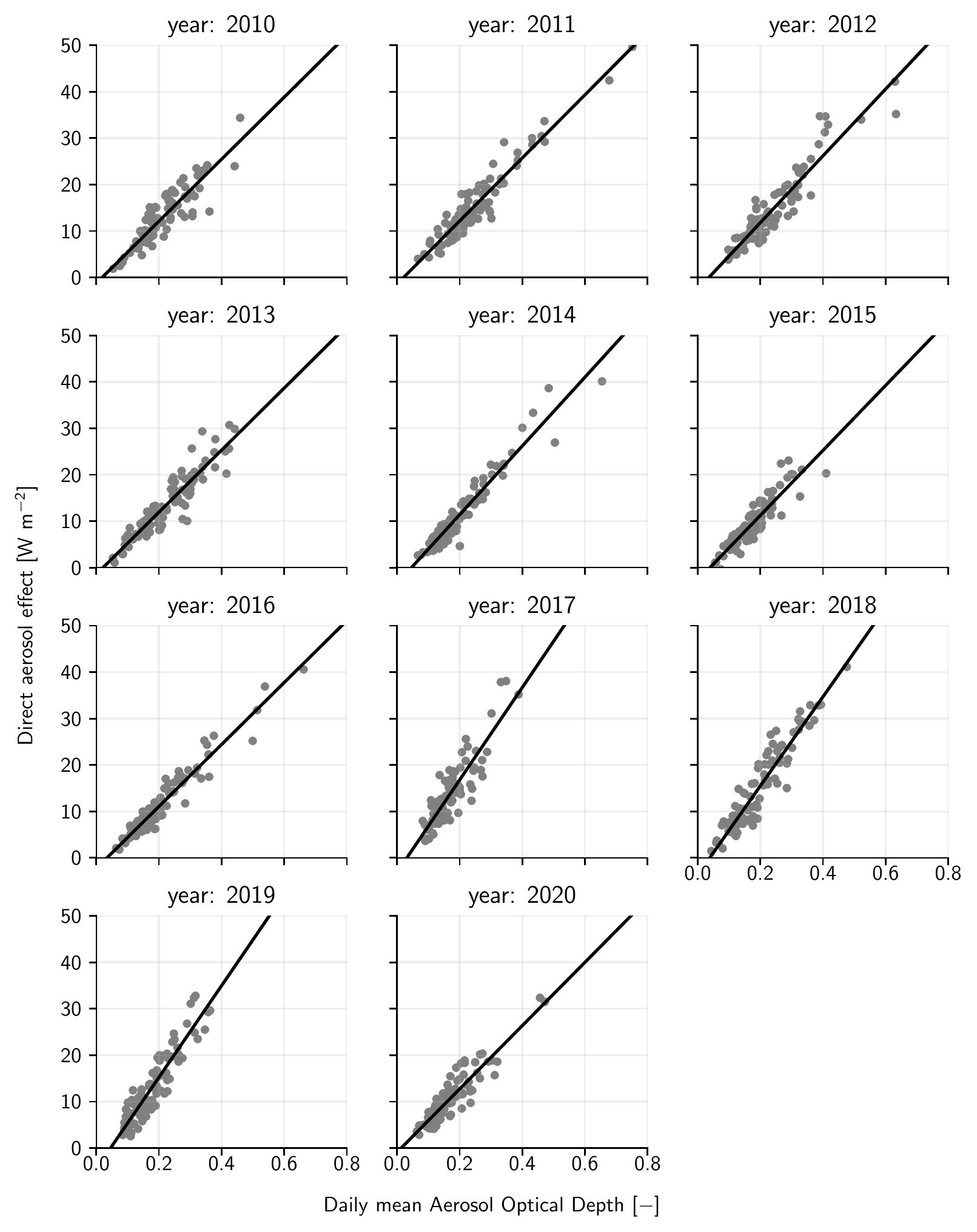}
\caption{Daily mean direct aerosol effect, computed as the difference between the clear-sky RTE+RRTMGP simulation with water vapor but without clouds and aerosols and the McClear clear-sky product, against daily mean aerosol optical depth. Each year is shown in a single panel and the data points represent all individual days in March, April, and May. The black lines indicate the linear trend.}
\label{fig:aod_validation}
\end{figure}

\begin{figure}[ht]
\centering
\includegraphics[width=0.9\textwidth]{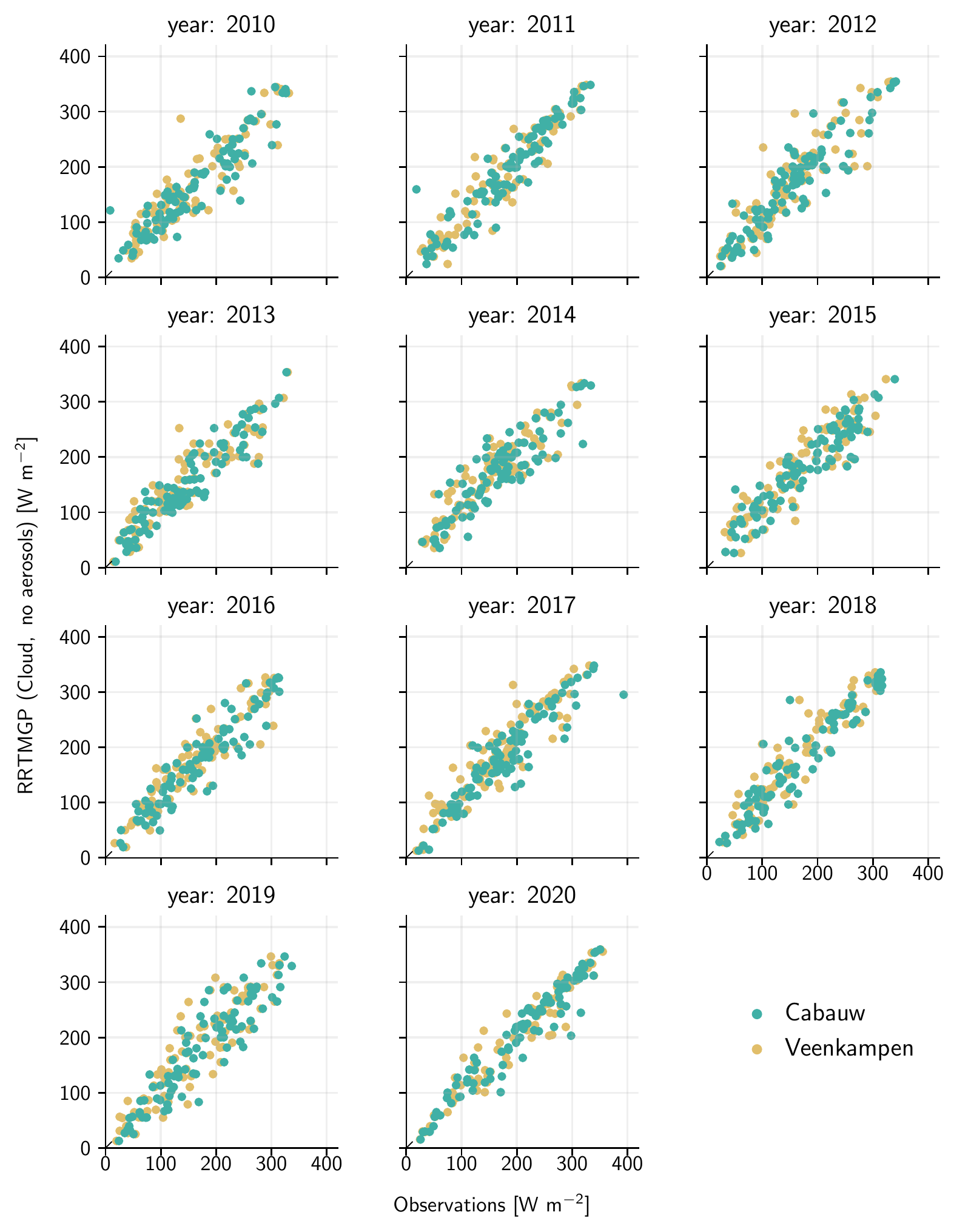}
\caption{Daily mean surface irradiances based on RTE+RRTMGP simulation with clouds but without aerosols against daily mean surface irradiances based on observations at Cabauw and at Veenkampen. Each year is shown in a single panel and the data points represent all individual days in March, April, and May.}
\label{fig:rrtmgp_validation}
\end{figure}

\begin{table}[ht]
\centering
\caption{Surface irradiance per experiment in units of W m$^{-2}$ for the period March, April, May for each of the years 2010--2020.}\label{tab:subl_rad_experiment}
\begin{tabular}{l|ccccccccccc}
                              & 2010  & 2011  & 2012  & 2013  & 2014  & 2015  & 2016  & 2017  & 2018  & 2019  & 2020  \\
\hline
Dry (\texttt{orange})         & 308.2 & 308.1 & 310.0 & 309.1 & 308.6 & 307.7 & 309.9 & 309.4 & 308.5 & 307.8 & 310.1 \\
Clear (\texttt{red})          & 264.6 & 262.4 & 263.4 & 265.3 & 261.9 & 263.8 & 264.8 & 262.9 & 261.4 & 262.6 & 266.2 \\
McClear (\texttt{light blue}) & 252.1 & 247.1 & 249.0 & 251.6 & 250.7 & 254.3 & 253.2 & 249.2 & 246.6 & 249.3 & 255.4 \\
Clouds (\texttt{dark blue}) & 170.5 & 191.9 & 168.5 & 153.4 & 173.5 & 179.6 & 173.6 & 182.1 & 179.9 & 174.2 & 208.3 \\
\hline
\end{tabular}
\end{table}

\end{document}